\documentclass{IEEEcsmag}

\graphicspath{{figs/}{figures/}{pictures/}{images/}{./}} 
\usepackage[colorlinks,urlcolor=blue,linkcolor=blue,citecolor=blue, hyperfootnotes=false]{hyperref}
\usepackage{booktabs}
\usepackage{balance}
\usepackage{fontawesome5}
\usepackage{xcolor}
\usepackage{xspace}
\usepackage{soul}

\usepackage{mdframed}

\newcounter{patterncounter}

\newcommand{\Pattern}{
  \medskip
  \stepcounter{patterncounter}%
  \noindent\faCubes\textsc{Pattern \#\thepatterncounter}: %
}

\newif\ifshowchanges\showchangesfalse
\newcommand{\inserted}[1]{%
    \ifshowchanges
        \textcolor{red}{#1}%
    \else
        #1%
    \fi}
\newcommand{\deleted}[1]{%
    \ifshowchanges
        \textcolor{gray!60}{{}#1}%

    \else\fi}

\newcommand{\Examples}{
  \smallskip
  \noindent\faFile*[regular]~\textsc{Examples}: %
}

\newcommand{\papertitle}{Agentic Visualization: Extracting Agent-based Design Patterns from Visualization Systems}

\jname{IEEE Computer Graphics \& Applications}
\jtitle{\papertitle}
\pubyear{2025}

\begin{document}


\sptitle{SPECIAL ISSUE}

\title{\papertitle}

\author{Vaishali Dhanoa}%
\affil{Aarhus University, Aarhus, Denmark and TU Wien, Vienna, Austria}

\author{Anton Wolter}%
\affil{Aarhus University, Aarhus, Denmark}

\author{Gabriela Molina Le{\'o}n}%
\affil{Aarhus University, Aarhus, Denmark}

\author{Hans-J{\"o}rg Schulz}%
\affil{Aarhus University, Aarhus, Denmark}

\author{Niklas Elmqvist}%
\affil{Aarhus University, Aarhus, Denmark}

\markboth{SPECIAL ISSUE}{SPECIAL ISSUE}

\begin{abstract}
    Autonomous agents powered by Large Language Models are transforming AI, creating an imperative for the visualization area. 
    However, our field's focus on a human in the sensemaking loop raises critical questions about autonomy, delegation, and coordination for such \textbf{agentic visualization} that preserve human agency while amplifying analytical capabilities.
    This paper addresses these questions by reinterpreting existing visualization systems with semi-automated or fully automatic AI components through an agentic lens.
    Based on this analysis, we extract a collection of design patterns for agentic visualization, including agentic roles, communication, and coordination.
    These patterns provide a foundation for future agentic visualization systems that effectively harness AI agents while maintaining human insight and control.
\end{abstract}


\maketitle
\begingroup
\renewcommand\thefootnote{}\footnotetext{\textit{Accepted for publication in IEEE Computer Graphics and Application. DOI: 10.1109/MCG.2025.3607741.}}
\addtocounter{footnote}{-1}
\endgroup


\chapteri{A}gents---software entities that perceive their environment and take goal-directed actions---have been a cornerstone of artificial intelligence research for decades~\cite{Wooldridge1995}.
The emergence of Large Language Models (LLMs) has dramatically accelerated this paradigm, enabling general-purpose agentic AI systems capable of sophisticated planning, reasoning, and autonomous action in complex domains~\cite{Park2023}~\cite{YE202443}.
Given how previous waves of AI innovation have transformed visualization, this agentic turn will inevitably reshape our field.
Yet visualization differs fundamentally from other AI application domains through its emphasis on human-centered reasoning, support for open-ended sensemaking, and placement of human analysts at the center of the analytical process.
The growing capabilities of agentic systems challenge the human-centered focus of visualization, raising questions about the level of agency, delegation, coordination, and responsibility between humans and AI~\cite{DBLP:journals/pnas/Heer19}.
What design patterns can guide us in developing agentic systems that leverage agentic capabilities while preserving human insight and analytical control?

To address these questions, we introduce the concept of \textit{agentic visualization} (AV): interactive visual analysis systems that incorporate autonomous agents(whether AI-driven or rule-based) while maintaining human agency in the analytical process.
Rather than starting from scratch, we analyze existing visualization systems, both old and new, with automatic or semi-automatic capabilities---sometimes described as mixed-initiative systems~\cite{DBLP:conf/chi/Horvitz99}---reinterpreting them as early forms of agentic visualization.
While new visualization systems use LLM-based agents, we intentionally focus on prior foundational literature (before the introduction of LLMs), to understand what we can learn from the past to inform our present design patterns for agentic visualization.

By examining these systems through an agentic lens, we identify recurring structures that effectively coordinate automated components and human analysts.
We formalize these structures as \textit{design patterns} for agentic visualization that balance automation and analytical control.
Design patterns, originating in architecture~\cite{Alexander1977} and software engineering~\cite{Gamma1994}, 
provide a structured vocabulary for describing reusable solutions to common challenges.
This approach aligns with human-centered AI principles that emphasize human values, interpretability, and agency in AI-augmented systems~\cite{Amershi2019, Shneiderman2022}.

We validate our proposed design patterns by identifying their presence across diverse published visualization systems.
Each pattern is structured with a descriptive name, the problem it addresses, its solution structure, implementation examples, and associated tradeoffs.
To demonstrate practical application, we present a novel example scenario that combines multiple patterns to create a dynamic visualization dashboard surveying an active scientific field where agents with different roles, communication, and coordination mechanisms work together to update progressively.

This paper makes the following contributions:
(1) The concept of agentic visualization as a framework for designing and analyzing systems that incorporate autonomous agents while preserving human-centered analytical processes;
(2) a collection of design patterns for agentic visualization extracted from a systematic analysis of prior foundational literature on automated and semi-automated visualization systems; and
(3) a demonstration showing how these patterns can be applied to a novel analytical scenario, including their composition into higher-order solutions.

\section{AGENTS AND AGENTIC VISUALIZATION}

Agents are software entities that are capable of perceiving their environment and acting upon it to achieve goals~\cite{Wooldridge1995}.
An agent's capacity to make their own decisions and take intentional actions towards that goals, is referred to as \textit{agency}, which lies on a spectrum from simple, reactive behavior to fully autonomous behavior.

We define \textit{agentic visualization} as the integration of agents with interactive visual analysis systems while balancing computational autonomy with human analytical control.
This approach enables agents to independently perform complex analytical tasks while preserving the human analyst's ability to guide the analytical process, interpret results, and maintain decision-making authority.

Agentic visualization extends and transforms several existing paradigms in visual analytics.
Building on \textit{mixed-initiative interaction}~\cite{DBLP:conf/chi/Horvitz99} and human-AI collaboration~\cite{Amershi2019}, agentic visualization specifically emphasizes agent autonomy beyond reactive assistance. Unlike traditional mixed-initiative systems~\cite{DBLP:journals/tvcg/UlmerAFKM24}, where humans and agents collaboratively initiate tasks, agentic visualization allows agents to proactively identify opportunities for analysis, suggest approaches, or even pursue analytical threads independently.

Instead of waiting for humans to issue commands and wait for response, agentic visualization challenges this traditional model by establishing a more fluid relationship where both human and agent can initiate analytical activities.
This shift prompts questions about how agents should behave when idle: should they remain passive until directed (\textit{cold start}), or should it proactively explore data and generate insights even without explicit user direction?
The answer likely depends on context, task complexity, and user preferences.

Agentic visualization also leverages progressive visual analytics~\cite{DBLP:journals/tvcg/UlmerAFKM24}, where systems provide incremental visual updates during computationally intensive processes.
However, while progressive analytics primarily focuses on making computation visible, agentic visualization extends this by enabling agents to make autonomous decisions about which computational paths to pursue based on intermediate results.
This way, agents can prioritize promising analytical directions, abandon unproductive paths, and adapt strategies without continuous human supervision.

To design such an agentic visualization system, we turn to both old and new visualization systems and reframe them through an agentic lens.
Through this process, we can identify patterns of effective human-agent collaboration that have emerged organically.

\begin{mdframed}[backgroundcolor=blue!10,
    frametitle={\section*{\textcolor{white}{Sidebar: Agents and AI}}},
    frametitlerule=true, frametitlebackgroundcolor=bgcolor]

Autonomous agents---software entities that perceive their environment and take goal-directed actions---have been a cornerstone of artificial intelligence research since the field's inception in the 1950s~\cite{Wooldridge1995}.
The early reactive agents of the 1990s emphasized real-time interaction over complex reasoning, exemplified by Rodney Brooks' subsumption architecture~\cite{DBLP:journals/ai/Brooks91}, followed by multi-agent systems and cognitive architectures that aimed to model distributed problem-solving and human cognition respectively~\cite{Laird2022}.

The current resurgence of agent technology, powered by Large Language Models (LLMs), has dramatically accelerated this paradigm, enabling general-purpose agentic AI systems capable of sophisticated planning, reasoning, and autonomous action in complex domains~\cite{Park2023}.
This represents a significant shift in capabilities but confronts persistent challenges that historical agent systems also faced: poor generalization across domains, scalability issues with complex tasks, inadequate coordination mechanisms, brittle performance in unexpected situations, and ethical concerns.
These limitations highlight an inherent tension in agent design that was famously debated by Ben Shneiderman and Pattie Maes at IUI '97 and CHI '97, where Shneiderman advocated for direct manipulation interfaces that ``affords the user control and predictability,'' while Maes argued for delegation to software agents that could learn user preferences and act on their behalf~\cite{DBLP:journals/interactions/ShneidermanM97}.
This debate continues to shape the balance between computational agency and human control in interactive systems.

Addressing these challenges requires more than just technical improvements to agent capabilities.
As Shah and White argue, successful agent systems need to generate sufficient value for users, provide adaptable personalization, establish trustworthiness, achieve social acceptability, and develop standardization protocols.
Aridor and Lange's work on design patterns for agent systems provides a useful foundation for capturing common solutions to recurring problems in agent design~\cite{DBLP:conf/agents/AridorL98}.
Their patterns are conceptually divided into three classes:

\begin{itemize}
    \item \textbf{Traveling patterns:} Agents moving and navigating in distributed environments;
    \item \textbf{Task patterns:} Accomplishing agent objectives and distributing workloads; and
    \item \textbf{Interaction patterns:} Governing communication and coordination between agents.
\end{itemize}

Recent work has extended this approach to foundation model-based agents, with researchers developing catalogs of architectural patterns that address challenges such as hallucinations, explainability, and accountability~\cite{DBLP:journals/jss/LiuLLZZXHW25}.

\end{mdframed}


\section{RELATED WORK}

The notion of agents lies on a spectrum of varying degrees of their inherent agency over their own actions.

On one end of that spectrum, agents are explicitly called upon by the user and given the autonomy to perform tasks under changing and unanticipated conditions without a user's oversight.
This ``autonomous systems'' approach is embodied in early visualization research dating back to the 1990s.
It can be found in concepts like \emph{Smart Particles} that are sent out into the data space, travel through it, and in their wake leave behind visual elements that make the data space visible as the swarm of particles wanders across it~\cite{Pang1994}.
Another example is the idea of Design Agents with different strategies and abilities and which by interacting with each other, yield visualizations as an emergent joint outcome that automatically adjusts to underlying data changes as the agents react to them~\cite{Ishizaki1996}.

The next step on that spectrum involves agents that operate continuously alongside the users without being explicitly triggered.
They silently observe the users' actions and offer unobtrusive help, even anticipating user goals by extrapolating their intent and performing likely tasks.
This echoes ideas such as Mark Weiser's \emph{Calm Computing}, sometimes likened to ``a raft of invisible servants''~\cite{Rogers2006}.
In visualization, we can find this idea in agent-based systems that run alongside or inside a visualization system and gently aid users, for example, in finding multi-view setups that maximize visibility and relevance of the shown data~\cite{Moreira2013} or by suggesting meaningful next steps in visual analyses~\cite{Sperrle2018}.

On the other end of the spectrum, these independent agents are proactive themselves and may even be permitted to interrupt the user if needed.
Visualization research has long explored such mixed-initiative interactions~\cite{DBLP:conf/chi/Horvitz99} where computational systems and human analysts divide the labor and collaboratively build understanding.
In progressive visualizations, for example, agents may monitor in-progress views for changes, alerting the users in case any relevant changes occur---in particular in off-screen views~\cite{Jo2021}.
This frees the human users from monitoring and allows them to explore the data freely without fearing to miss out on anything happening elsewhere.
Similarly, independent processes can be used to check a visualization design for construction errors and readability issues (\emph{visualization linters}), or check an ongoing visual analysis for potential biases---again alerting users if any issues are detected.
\section{PATTERNS FOR AGENTIC VISUALIZATION}

\begin{figure*}[htb]
     \centering
     \includegraphics[width=\textwidth, alt={Agentic Visualization Patterns.}]{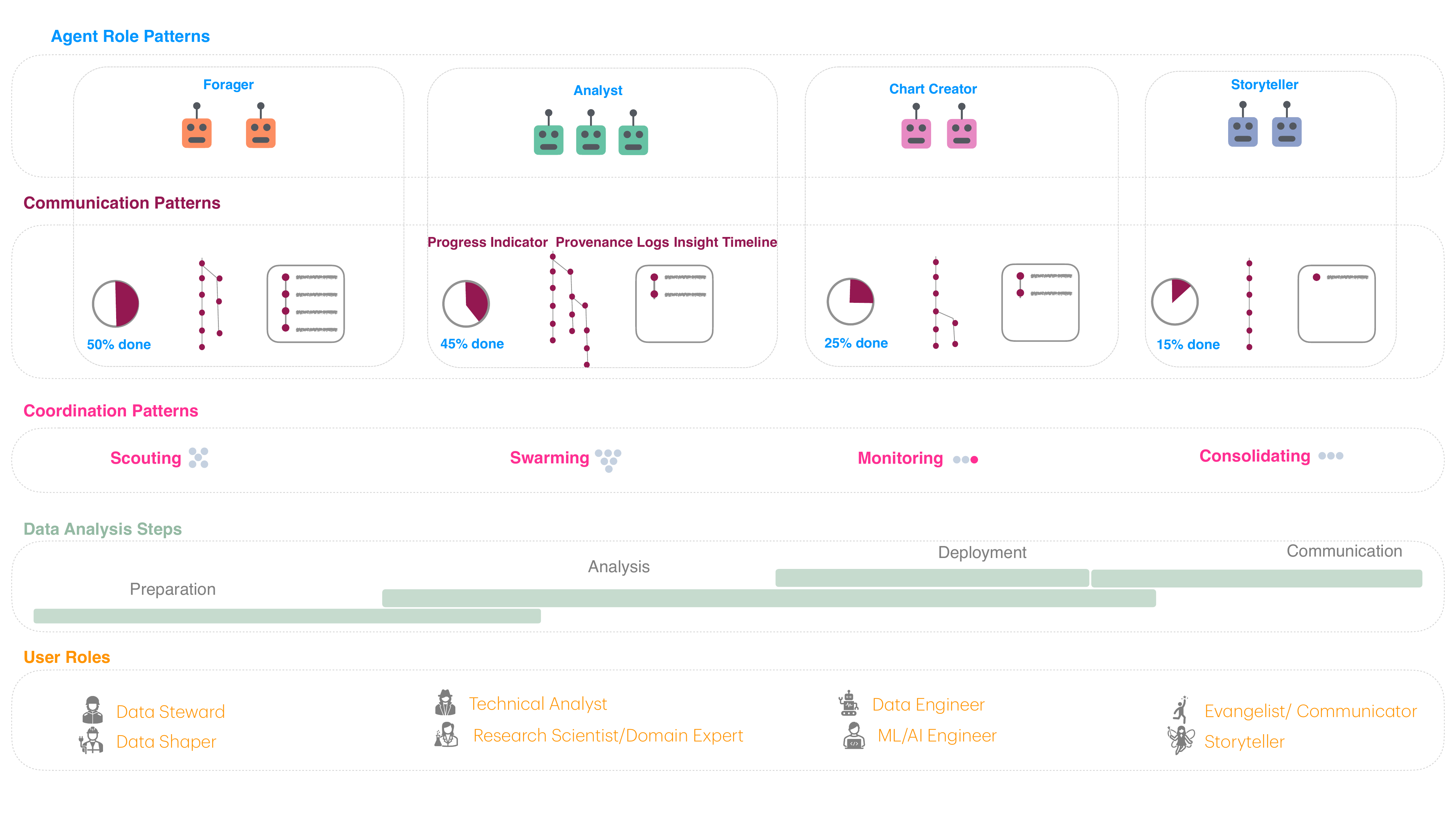}
     \caption{\textbf{Overview of agentic visualization patterns.}
     Agent role patterns running their computations throughout the data analysis workflow and interacting with various human roles, based on Crisan et al.~\cite{crisan2021}.}
     \label{fig:agentic_visualization_patterns}
\end{figure*} 

To systematically identify patterns for agentic visualization, we conducted a comprehensive analysis of visualization literature where autonomous or semi-autonomous components collaborate with humans and among themselves.
Our methodology focused on extracting recurring design solutions that effectively balance computational agency with human control in visual analytics systems.

Our review focused on prior work that met four specific inclusion criteria: (1) involves direct human user interaction; (2) has an emphasis on data and data analysis; (3) includes interactive visualization components; and (4) contains autonomous or semi-autonomous computational elements.
We did not limit our analysis to systems explicitly labeled as ``agent-based,'' recognizing that many visualization systems incorporate autonomous behaviors without using this terminology.
Instead, we focused on functional characteristics that exemplify human-agent collaboration in visual analytics contexts.
Furthermore, we restricted our search to academic publications for this survey. 

From this analysis, we identified patterns that appeared in at least two distinct systems, suggesting their broader applicability.
We also selected patterns that, while perhaps less common in the literature, show particular promise for agentic visualization research.

Following long-standing design pattern praxis~\cite{Alexander1977, Gamma1994}, each pattern in our catalog has a descriptive \textbf{title} capturing the pattern's essence; the specific \textbf{problem} the pattern addresses; the reusable \textbf{solution} that resolves the problem; and the \textbf{consequences} of implementing the pattern, including benefits, limitations, and tradeoffs.
Our analysis revealed three distinct categories of design patterns for agentic visualization:

\begin{enumerate}
    \item \textbf{Agent Role Patterns:} These patterns define the fundamental roles and responsibilities agents can assume within a visualization system.
    They establish the possible scopes of agent autonomy and the division of analytical labor between human and computational actors.
    
    \item \textbf{Communication Patterns:} These patterns structure the information exchange between agents and humans.
    They address how agents present findings, receive feedback, and maintain transparency about their operations and reasoning processes.

    \item \textbf{Coordination Patterns:} These patterns govern the temporal and workflow dynamics between agent and user activities. 
    They establish when and how agents act, how concurrent human-agent actions are synchronized, and how analytical control shifts between participants.

\end{enumerate}

Table~\ref{tab:pattern-overview} presents our proposed catalog of agentic visualization patterns organized by category. 
This catalog is not exhaustive but provides a foundation for understanding and designing effective agentic visualization systems. 
In the following sections, we elaborate on each pattern category and detail the specific patterns within them, providing concrete examples from existing systems and discussing their implications for future research and development in agentic visualization.

\newcommand{\PatternTimeline}{Insight Timeline\xspace}
\newcommand{\PatternForager}{Forager\xspace}
\newcommand{\PatternAnalyst}{Analyst\xspace}
\newcommand{\PatternChartCreator}{Chart Creator\xspace}
\newcommand{\PatternStoryteller}{Storyteller\xspace}
\newcommand{\PatternProgress}{Progress Indicator\xspace}
\newcommand{\PatternProvenance}{Provenance Log\xspace}
\newcommand{\PatternSwarming}{Swarming\xspace}
\newcommand{\PatternConsolidating}{Consolidating\xspace}
\newcommand{\PatternScouting}{Scouting\xspace}
\newcommand{\PatternMonitoring}{Monitoring\xspace}

\begin{table}[htb]
    \centering
    \begin{tabular}{m{7em}ll}
    
        \toprule
        \textbf{AV Category} & \textbf{Pattern} & \textbf{Examples}\\
        \toprule
        
        Role Patterns 
        & \PatternForager & \cite{gallimore99, vande07}\\ 
        & \PatternAnalyst & \cite{srinivasan21snowy, wang2025jupybara}\\ 
        & \PatternChartCreator & \cite{dataformulator2024, liu24ava}\\ 
        & \PatternStoryteller & \cite{fu2025dataweaver, shen24story}\\ 
        
        \midrule
        
        Communication Patterns
        & \PatternTimeline & \cite{badam2017steering, cui19datasite}\\
        & \PatternProgress & \cite{badam2017steering, Jo2021}\\
        & \PatternProvenance & \cite{DBLP:conf/sigmod/CallahanFSSSV06, park2022storyfacets}\\
        
        \midrule
        
        Coordination Patterns
        & \PatternScouting & \cite{Pang1994, cui19datasite} \\
        & \PatternSwarming & \cite{dynamicdata2010, Moere2008} \\
        & \PatternMonitoring & \cite{badam2017steering, Jo2021} \\
        & \PatternConsolidating & \cite{insightlens2025, shen24story} \\
        
        \bottomrule
    \end{tabular}
    \smallskip
    \caption{\textbf{Overview of our agentic visualization patterns.}
    The identified patterns and their corresponding categories.
    For each pattern, we give several examples from existing visualization systems that showcase the pattern.}
    \label{tab:pattern-overview}
\end{table}

\begin{mdframed}[backgroundcolor=blue!10,
    frametitle={\section*{\textcolor{white}{Sidebar: Visualization and Agents}}},
    frametitlerule=true, frametitlebackgroundcolor=bgcolor]

While agents have not been a major theme in the scientific community for visualization so far, there have been several instances of autonomous and semi-autonomous entities in visualization systems in the literature.
Here, we review these prior systems and their autonomous components. 

\begin{itemize}
    
    \item \textbf{InsightsFeed}~\cite{badam2017steering} is a progressive visual analytics system that generates and continuously refines an analysis of social media posts based on keywords, sentiment, and multidimensional embedding.

    \item \textbf{VisTrails}~\cite{DBLP:conf/sigmod/CallahanFSSSV06} manages visualization workflows and their provenance metadata, enabling reproducibility, comparison of visualization pipelines, and systematic tracking of the scientific discovery process.

    \item \textbf{DataSite}~\cite{cui19datasite} is a proactive visual analytics system that autonomously executes background computations to identify salient data features and surfaces these findings as notifications in a feed timeline.

    \item \textbf{\textsc{DataWeaver}}~\cite{fu2025dataweaver} is a data story authoring system that supports \textit{vis-to-text} and \textit{text-to-vis} compositions through statistics, custom algorithms, and LLMs. 

    \item \textbf{SurfaceMapper}~\cite{gallimore99} identifies relevant surfaces in 3D scientific data through multiple complementary agents that perform independent data interpretations. 
    
    \item \textbf{HaLLMark}~\cite{hoque2023hallmark} is a writing assistance system that autonomously captures and visualizes the complete provenance of writer-LLM interactions.

    \item \textbf{ProReveal}~\cite{Jo2021} is a progressive visual analytics system that autonomously monitors intermediate analytical findings and activates safeguards when uncertain knowledge requires verification.

    \item \textbf{AVA}~\cite{liu24ava} is a visualization agent framework that uses multimodal LLMs to autonomously interpret visual outputs and make visualization parameter decisions.

    \item \textbf{\textsc{Data Formulator}}~\cite{dataformulator2024} is an interactive visualization authoring tool that uses an AI agent to create a visualization from raw data based on user's intent provided as a data concept.

    \item \textbf{StoryFacets}~\cite{park2022storyfacets} is a collaborative analysis and storytelling system that autonomously maintains visual provenance across multiple linked views. 
    
    \item \textbf{Data Director}~\cite{shen24story} is a multi-agent system where LLM-powered agents generate insights, visualizations, and text to integrate them as a story into animated data videos.

    \item \textbf{Snowy}~\cite{srinivasan21snowy} is a natural language interface that generates contextual utterance recommendations by combining data interestingness metrics and language pragmatics to guide analytic workflows. 

    \item \textbf{\textsc{DynaVis}}~\cite{vaithilingam24} is a multimodal system that combines a natural language interface with WIMP elements to generate standard charts and provide automatically synthesized widgets for visualization editing.

    \item \textbf{In-formation flocking}~\cite{vande07} is a visualization approach that leverages a decentralized multi-agent system to recognize dynamic patterns and clusters in time-varying datasets, with local and global views.

    \item \textbf{InsightLens}~\cite{insightlens2025} is an LLM-agent-based interactive system that automatically captures and manages insights from users' conversational data analysis workflow.

 \item \textbf{Jupybara}~\cite{wang2025jupybara} is an LLM-multi-agent-based Jupyter Notebook extension that supports analysts during exploratory data analysis and storytelling via a natural language interface.
    
\end{itemize}


\end{mdframed}

\section{AV ROLE PATTERNS}

We identified four main agent roles for visualization.
Each pattern is designed to support human-agent collaboration, where the human may have varying levels and types of expertise. We highlight relevant human roles for each pattern using the data science role described by Crisan et al.~\cite{crisan2021}, as illustrated in Figure~\ref{fig:agentic_visualization_patterns}. These roles help inform how users engage with agents, but our primary focus remains on the agent's roles and their behavior.

\Pattern
The \textbf{\PatternForager} pattern describes agents that search for relevant data based on parameters set by humans, typically data shapers or analysts. This pattern deals with the challenges of processing large datasets in parallel, while adjusting to the domain context.
\PatternForager agents often operate in groups, each applying custom rules or parameter variations to explore data in parallel and report their findings.
\Examples
In SurfaceMapper~\cite{gallimore99}, multiple agents with complementary expertise cooperate to identify data of interest in large volumes of scientific data.
The agents report their findings to the human user and exchange them to revise their hypotheses. 
The in-formation flocking~\cite{vande07} technique involves multiple \PatternForager agents that follow a set of behavior rules with the goal of highlighting potential patterns of interest in time-varying datasets.
Each agent focuses on specific data neighborhoods and calculates similarity metrics to detect temporal changes.

\PatternForager reduces the manual effort of data shapers and analysts by collecting samples of interest and providing preliminary insights that these users can compare and synthesize. Due to their flexibility, these agents can adapt across domains, enabling domain experts to specify their needs and configure the agents accordingly.

\Pattern
An \textbf{\PatternAnalyst} agent generates and guides human users, typically a data analyst, in making sense of a dataset. 
This pattern is associated with the challenges of exploring any data without specific goals to steer the process.
To assist people who may struggle to choose a starting point for their exploration (i.e., the \textit{cold start} problem), \PatternAnalyst agents recommend multiple ways to start.
The agent analyzes the dataset and prepares an initial set of recommendations.
As soon as the person starts to interact, the agent starts learning from those actions to make follow-up recommendations.
The \PatternAnalyst encourages breadth-oriented exploration to increase data and analytical coverage.

\Examples
Snowy~\cite{srinivasan21snowy} generates multiple natural language inquiries based on patterns detected automatically in the dataset.
Jupybara~\cite{wang2025jupybara}, an LLM-based multi-agent AI assistant, helps the data analyst as they explore their data in a Jupyter Notebook.

With the support of a \PatternAnalyst, users worry less about missing key patterns or unexplored areas. However, frequent recommendations can become overwhelming and hinder focus. Jupybara~\cite{wang2025jupybara} addresses this by displaying limited details for clarity. Future work could build on this to reduce distraction and provide clear information to the user.

\Pattern
\textbf{\PatternChartCreator} agents assist users, such as technical analysts, domain experts, and evangelists, who regularly create and refine visualizations as part of their workflows. This pattern applies to scenarios where the goal is to design one or more visualizations, individually or as part of a composite view, such as a dashboard. This process involves a series of decisions, such as choosing an appropriate visual encoding and adjusting visualization parameters. These decisions often require multiple iterations and a combination of visualization literacy and technical skills, depending on the tool. 
\PatternChartCreator agents help reduce the workload by helping the authoring process.

\Examples
In \textsc{DynaVis}~\cite{vaithilingam24}, the agent creates charts on demand and generates editing widgets when the person asks to change a visualization parameter.
Data Formulator~\cite{dataformulator2024} uses an agent to transform raw data to visualizations based on author-defined data concepts.
One challenge of this pattern is that interacting with the agent is often the only way to create visualizations, potentially adding an extra layer of complexity for simple changes.
Moreover, the automatically created elements are constrained to specific visualizations and widgets. 
Still, co-creating with an agent can reduce the need for repetitive tasks and advanced knowledge.

\Pattern
\textbf{\PatternStoryteller} pattern provides agents that support the creation and refinement of data-driven stories by combining text, insights, and visual elements into a cohesive narrative. These agents complement the work of user such as evangelists, analysts, and domain experts, by taking over the time-intensive tasks such as text writing, visualization authoring, video creation, and adding annotations.

\Examples
\textsc{DataWeaver}~\cite{fu2025dataweaver} generates narrative text through dedicated algorithms and LLMs and integrates them into the main story.
Data Director \cite{shen24story} provides an agent for designing and synchronizing visual animations for a data video.
Delegating storytelling tasks to agents allows users to focus on specific story elements while offloading the generation of others. This can speed up the process and lower barriers for users who may not have the necessary expertise. However, this delegation introduces a tradeoff, where the agent's output must still be reviewed for accuracy, which can be challenging without sufficient knowledge to evaluate it.

\section{AV COMMUNICATION PATTERNS}

Agents deployed for addressing visualization-specific tasks---from analysis to storytelling---should also communicate and present their findings to the user.
Here we discuss these \textbf{communication patterns} through which agents and users share task allocation, progress, and completion.

\Pattern
The \textbf{\PatternTimeline} pattern deals with situations where humans and agents need a way to synchronize and share findings while continuing to work autonomously on their own tasks.
The solution is to provide a common channel for them to share new insights and findings collected during the data analysis process.
An Insight Timeline serves as a channel where both ongoing and explored tasks can be shared to (i)~declare the current focus of exploration within the dataset to avoid redundancy, (ii)~report interesting findings, and (iii)~steer the analysis direction.

\Examples
InsightsFeed~\cite{badam2017steering} explores large textual datasets and progressively updates the results in an \PatternTimeline so that the user, such as analysts and domain experts, can steer their analysis direction based on the results shared by the system.
Similarly, DataSite~\cite{cui19datasite} adopts the UI principles from InsightsFeed to propagate the results of background computation to the user without interrupting the cognitive process of the user.
Users can also rank, sort, and filter the insights based on their needs.


\Pattern
The \textbf{\PatternProgress} pattern supports communication about which tasks the agents have undertaken and how far along each task is.
This pattern helps users, such as analysts, engineers, and domain experts, interpret intermediate results, assess uncertainty in the current analysis, and identify areas of interest.

\Examples
InsightsFeed~\cite{badam2017steering} shows the progress and aliveness by using a blinking rectangle next to a progress bar to highlight the current chunk of data being processed by the system.
The system also uses a quality/stability metric to indicate the quality or uncertainty of the current state with a small line chart.

ProReveal~\cite{Jo2021} shows the progression of both ongoing and completed visualizations to the analyst using a progress ring, which is hidden once the visualization is completely rendered with all the data.
In both InsightsFeed and ProReveal, users can interact with this process through pause, resume, and remove buttons to control the progression and explore the data in its current state.
The progress indicators can also be used for other measures such as time spent vs.\ total time left, the amount of data processed, and the speed of the task performance.

\Pattern
The complex and collaborative data analysis process makes it difficult to track what steps were performed, by whom and when.
To ensure reproducibility, verification, and replication of analysis steps, the \textbf{\PatternProvenance} pattern provides a common audit log of sources, transformations, and filters applied to the data by both human and AI agents.

\Examples
VisTrails~\cite{DBLP:conf/sigmod/CallahanFSSSV06} captures provenance for visualization by tracking all the changes that users, especially data shapers and analysts, make to the visualization pipeline in an exploration task.
Users can interact with the provenance information---presented as a directed graph---to reproduce a specific state and understand the process history.
For collaborative exploration, VisTrails allows users to integrate other analysts' workflows afterward in their view to analyze the provenance information.
StoryFacets~\cite{park2022storyfacets} presents a simpler view for displaying provenance from multiple collaborators on the same analysis session by creating multiple linked formats called facets. HaLLMark~\cite{hoque2023hallmark} uses summary statistics to encode the percentage of tasks performed by human and AI in a collaborative writing process.



\section{AV COORDINATION PATTERNS}

Agentic visualization workflows often employ multiple agents.
Depending on the nature of the task, agents may work towards a shared goal, operate independently, or wait for others to complete their tasks before proceeding. 
Coordination is essential to ensure that agents and humans can schedule and execute their tasks effectively.

Here we discuss different coordination patterns that describe how agents and users coordinate during analysis. 
 
\Pattern
At the beginning of the analysis process, it is often unclear which tasks should be prioritized to perform the analysis effectively, especially in large datasets.
\textbf{\PatternScouting} can serve as an initial coordination pattern, where multiple autonomous agents independently explore the dataset to identify potentially interesting patterns, anomalies, or other areas of interest.
Users, typically data shapers, stewards, or domain experts, can provide a single or multiple goals, and the agents can operate with this knowledge and adapt to their feedback.

\Examples
Mix\&Match~\cite{Pang1994} releases smart particles (i.e., agents) into the dataset to highlight features of interest. 
The particles operate on rules, have a finite lifetime, and scout for information in the dataset, leaving behind visual effects for the users.
Users can also modify particle behavior by updating the rules to produce different effects.
DataSite~\cite{cui19datasite} proactively runs background computations on datasets to support users during the different stages of the visual data exploration process.
The user still maintains agency over the exploration and is actively supported by the system to generate insights, creating a balanced dialogue between the analyst and the system (or agents) and utilizing their respective strengths.

\Pattern
In some cases, it might not be enough to employ autonomous agents on independent tasks with different goals, but multiple agents should be deployed with a single task and a shared goal.

\textbf{\PatternSwarming} agents can address this issue by dividing the problem into smaller components that each agent in the swarm can autonomously work on and coordinate. 

\Examples
To solve a clustering problem, Saka et al.~\cite{dynamicdata2010} designed a simultaneous clustering and visualization algorithm that uses an ongoing swarm of agents to coordinate the tasks with each other. 
They use a flock-based approach for data clustering where each agent receives a position in a 2D or 3D visualization panel.
Similar agents come together to form swarms to visualize dynamic clustering where data is collected over time.
Vande Moere~\cite{Moere2008} proposed visualization agents that determine their own visual properties through continuous interactions with other agents.
Their iterative organization creates visual effects based on data properties that both agents and users can perceive.
These agents use the concept of \PatternSwarming and can identify other agents in their vicinity, read their data and move towards or away from them based on their data properties.
Using a stock dataset, where agents represent companies, Moere describes how agents flock together or apart based on the similarity of their price change behavior over time, leading to different patterns such as short-term clustering and long-term zoning patterns.


\Pattern
\textbf{\PatternMonitoring} enables agents launched during the analysis process to track temporally streaming data and notify the users, such as analysts, domain experts, among others, of any changes, without overwhelming them. 

\Examples
InsightsFeed~\cite{badam2017steering} enables users to keep track of evolving results during progressive data analysis.
In addition to the visual exploration of incrementally processed data, users can also monitor the progress of the background computation and see how far along it is.
DataSite~\cite{cui19datasite} also uses background processes to support the users in visual data exploration and help them actively monitor the results of these processes using a feed timeline.
In ProReveal~\cite{Jo2021}, users can start the exploration and employ agents with different analytical support to guard the analysis process as a large dataset is progressively explored. It uses a web-based user interface where users can explore the dataset, employ the guards, and continue exploring while \PatternMonitoring the guards' progress.
They are notified when a change occurs and the condition is modified.

\Pattern
As multiple agents perform various tasks throughout the analysis process, \textbf{\PatternConsolidating} the outcomes becomes essential to help users, such as evangelist, storyteller and agents to build upon these findings. 
Based on the consolidated results, further exploration, summarization, or sensemaking of the data can be done.
For example, after \PatternConsolidating the findings from \PatternScouting the dataset, users may wish to deepen their exploration in specific areas of interest.
This is especially useful during sensemaking, where users can interpret and communicate important findings from consolidated information.

\Examples
InsightLens~\cite{insightlens2025} uses an LLM-agent-based framework that automatically records and organizes insights during conversational data analysis with a human and an LLM agent.
Based on the outcome of the analysis, the user can monitor and navigate the conversation using visualizations and receive guidance on further data exploration without interrupting their workflow.
DataDirector~\cite{shen24story} uses multiple agents to automate the creation of animated videos from data.
LLM-based agents autonomously interpret the data, analyze it and further spawn design agent roles to create final data videos.
In this case, tasks are broken down into sub-tasks at each analysis step, and the results from these tasks serve as input to the next step.
In the final step, the main controller agent consolidates the decisions made by the agents to generate a data video for the users.
\section{APPLYING AV PATTERNS}

To demonstrate the practical application of our proposed AV patterns, we present a \textbf{Medical Research Monitoring} scenario that addresses the challenge of keeping pace with rapidly expanding medical literature.
Consider a research team in a medical domain, where thousands of new publications emerge continuously across various sources that may be relevant to the field of study or a certain treatment or medication. 
The volume and velocity of this research make comprehensive manual monitoring impossible, yet staying current is critical.

Traditionally, identifying key papers and extracting meaningful information to visualize the literature landscape and synthesize findings requires extensive manual effort.
In this scenario, the primary user roles include a research scientist, a technical analyst, and an evangelist.
Not all roles from the broader data science framework~\cite{crisan2021} are present, reflecting the current structure of the organization.
This complex scenario leads to several interconnected challenges:

\faSearch~\textbf{Challenge 1: Query formulation and seeding.}
Crafting effective search strategies that capture relevant publications requires domain expertise, often from research scientists.

The \PatternForager pattern can support researchers by deploying multiple autonomous agents that can explore different search strategies simultaneously.
Each \PatternForager agent can be configured with complementary expertise, and these agents can autonomously refine their search parameters based on initial results.
Additionally, the \PatternAnalyst pattern can assist research scientists by analyzing existing literature collections to suggest relevant search terms and identify potential gaps in current search strategies.

\faSlidersH~\textbf{Challenge 2: Scale and filtering.}
Research scientists and technical analysts need to manage the overwhelming volume of daily publications by applying efficient filtering techniques to identify relevant studies without missing critical information.
The \PatternScouting coordination pattern provides a solution for handling the massive scale of incoming publications. 
Multiple scouting agents can be deployed to independently explore different publication streams, with each applying domain-specific filtering criteria while maintaining awareness of overall research priorities.
This approach, combined with \PatternMonitoring, enables researchers and analysts to keep track of the progress of filtering operations across multiple databases simultaneously.
The \PatternProgress communication pattern keeps researchers informed about processing status without overwhelming them with intermediate results, allowing them to focus on high-priority publications while background agents handle the volume.

\faLightbulb~\textbf{Challenge 3: Concept extraction and validation.}
For technical analysts, automatically identifying and extracting key findings, methodologies, and outcomes from diverse publication formats is a complex task that inherently involves close collaboration with scientists to ensure accuracy and uphold reporting standards.
The \PatternAnalyst pattern excels at automated concept extraction, using natural language processing and domain-specific knowledge to identify key findings across diverse publication formats.
These agents can generate structured summaries of extracted concepts and present them through the \PatternTimeline communication pattern.
Additionally, \PatternChartCreator agents can generate domain-specific visualizations such as temporal trends, citation networks, and medication-specific charts that highlight key elements and emerging topics.
The \PatternProvenance pattern ensures that all extracted concepts maintain links to their source publications, enabling analysts and researchers to verify accuracy and trace findings back to original sources for validation.

\faProjectDiagram~\textbf{Challenge 4: Research cluster identification.}
The research scientists are responsible for discovering complex relationships between studies, including methodological similarities, conflicting results, and emerging research directions. This requires sophisticated analysis of both explicit citations and implicit conceptual connections.
The \PatternSwarming coordination pattern is particularly well-suited for discovering complex relationships between studies.
Multiple agents can work collaboratively to identify clusters based on different similarity metrics---methodological approaches, research populations, outcome measures, or citation networks. 
Each agent in the swarm contributes its specialized perspective while coordinating with others to build an understanding of research relationships. 
The \PatternTimeline pattern facilitates sharing of discovered clusters among team members, enabling collaborative validation and refinement of identified data patterns.

\faLayerGroup~\textbf{Challenge 5: Knowledge synthesis and gap analysis.}
The complex process of integrating findings across studies to maintain a coherent overview of the research landscape and systematically identify opportunities for future research while managing contradictory evidence and varying quality levels presents ongoing challenges.
The \PatternStoryteller pattern addresses the complex task of synthesizing knowledge across multiple studies. 
These agents can generate coherent narratives that integrate findings from related research clusters, identify contradictory results, and highlight emerging trends. 
The \PatternConsolidating coordination pattern ensures that insights from multiple agents are systematically integrated into comprehensive research overviews. 
Combined with \PatternProvenance, researchers can trace the synthesis process and understand how specific conclusions were derived from the underlying evidence.

\faChartLine~\textbf{Challenge 6: Temporal monitoring and adaptation.}
The ongoing need for continuously updating the monitoring system as new research emerges and research priorities evolve requires adaptive strategies that can respond to shifting focus areas and changing research methodologies.
The \PatternMonitoring coordination pattern provides continuous oversight of the evolving research landscape. 
Monitoring agents track changes in publication patterns, emerging topics, and shifting research priorities, alerting researchers and evangelists when significant developments occur. 
The \PatternProgress pattern communicates the evolution of research areas over time, while the \PatternTimeline serves as a living record of research developments.
This temporal awareness enables the system to adapt its focus areas dynamically, ensuring that monitoring strategies remain aligned with current research needs and emerging opportunities.
\section{RESEARCH VISION AND OUTLOOK}

Agentic visualization systems must navigate the relationship between computational autonomy and human analytical control.
As Shneiderman~\cite{Shneiderman2022} points out, human versus AI control is not a zero-sum game, but can be thought of as a 2D design space where human and computer control are independent axes.
This framing reveals opportunities where both humans and agents can retain high degrees of control.

We are currently witnessing the rise of agents throughout the visual analysis process, and this is expected to grow.
With their computational capabilities, agents help address scalability challenges and reduce the time from data to insight. Through our extracted design patterns and use case scenario, \deleted{multiple agents will play a collaborative role by supporting diverse human users by taking over cognitively demanding or repetitive tasks.}
\inserted{we see role of visualizations as two-fold: as an interpreter during agent autonomous task operation and as a channel for presenting final analysis outcomes.}

Several commercial tools, such as Microsoft Power BI and Tableau, have started integrating agents that support users from data processing to communication. While promising, the performance, utility and accuracy of these agents require further study to assess how effectively they amplify human capabilities.

Depending on the level of agency, the agents may only react in response to user input, such as in the case of Jupybara~\cite{wang2025jupybara}, Data Formulator~\cite{dataformulator2024}, \inserted{WaitGPT~\cite{xie2024waitgpt}, and LightVA~\cite{zhao2024lightva},} or proactively monitor user behavior to make suggestions, such as in the case of Ava~\cite{liu24ava} and InsightLens~\cite{insightlens2025}.
In reactive systems, like \inserted{LightVA~\cite{zhao2024lightva} and Jupybara~\cite{wang2025jupybara}, agents generate and plan tasks based on user input.} 
\deleted{Jupybara~\cite{wang2025jupybara}, a multi-agent system responds with an ordered plan to fulfill complex user queries.}
\inserted{In case of WaitGPT~\cite{xie2024waitgpt} and LightVA~\cite{zhao2024lightva}, users can also monitor and control the analysis process using flow diagrams and \PatternProgress.}
\deleted{To ensure full transparency, it is essential to provide additional communication information, such as \PatternProvenance and \PatternProgress, especially with large datasets where processing might take longer.
In multi-agent systems, coordination patterns also play a critical role.}

In proactive scenarios, agents may provide background and contextual information for insights, make suggestions for analytical directions, and provide careful provenance and source information for all decisions while leaving ultimate decision-making and sensemaking to the human analyst.

Practically speaking, we suggest deploying agentic visualization for the lower parts of the sensemaking loop~\cite{DBLP:conf/chi/RussellSPC93}---information foraging tasks at lower abstraction levels that involve data processing, pattern detection, and routine analytical operations.
The higher, more cognitively demanding levels can then be reserved for human analysts: forming hypotheses, making contextual judgments, and drawing conclusions that require domain expertise and ethical reasoning.

The shift toward agentic visualization raises questions about analytical responsibility, interpretability, and the preservation of human values that have long defined our field.
When an agent discovers a pattern that influences policy decisions, questions of responsibility become complex.
How do we maintain transparency when agents make autonomous analytical choices? What decisions should agents be allowed to make?
How do we ensure that increased automation enhances rather than undermines human analytical reasoning?
These concerns extend beyond traditional AI ethics frameworks.

The patterns presented in this work may be helpful to address these issues. 
For example, \PatternProvenance can maintain comprehensive logs of data sources and transformations applied to the data by both agents and users to ensure transparency and reproducibility of the analysis process. 
Still, ultimately, domain experts are required to validate the quality of the results.

Visualization has always emphasized human insight and interpretability as core values; put differently, the ability to ``use vision to think''~\cite{Card1999} is fundamental to our area.
Accordingly, agentic systems must preserve these values while expanding analytical capabilities.
This requires agents that can explain their analytical choices, expose their reasoning processes, and maintain clear audit trails of autonomous actions.
As analysts, we will still need to be able to think through the evidence provided to our eyes.
The patterns and principles outlined in this paper aim to guide us in achieving this balance.









\section*{Acknowledgments}

This work was supported partly by Villum Investigator grant VL-54492 by Villum Fonden.
Any opinions, findings, and conclusions expressed in this material are those of the authors and do not necessarily reflect the views of the funding agency.

\def\refname{REFERENCES}



\begin{IEEEbiography}{Vaishali Dhanoa} is a postdoctoral researcher in the Department of Computer Science at Aarhus University in Aarhus, Denmark.
Her research interests include data visualization, dashboard onboarding, storytelling, and human-centered AI.
Dhanoa received the Ph.D.\ degree in 2024 in computer science from Johannes Kepler University Linz.
Contact her at dhanoa@cs.au.dk.
\end{IEEEbiography}

\begin{IEEEbiography}{Anton Wolter} is a Ph.D.\ student in the Department of Computer Science at Aarhus University in Aarhus, Denmark.
His research interests include data visualization, agentic AI, and software engineering.
Wolter received a Master's degree in 2021 in computer science from NORDAKADEMIE.
He is a student member of the IEEE.
Contact him at wol@cs.au.dk.
\end{IEEEbiography}

\begin{IEEEbiography}{Gabriela Molina Le{\'o}n} is a postdoctoral researcher in the Department of Computer Science at Aarhus University in Aarhus, Denmark.
Her research interests include data visualization, human-computer interaction, and computer-supported cooperative work.
Le{\'o}n received the Ph.D.\ degree in 2024 in computer science from the University of Bremen.
Contact her at leon@cs.au.dk.
\end{IEEEbiography}

\begin{IEEEbiography}{Hans-J{\"o}rg Schulz} is an associate professor in the Department of Computer Science at Aarhus University in Aarhus, Denmark.
His research interests include information visualization, visual analytics, and human-data interaction.
Schulz received his doctorate degree in 2010 in computer science from the University of Rostock.
Contact him at hjschulz@cs.au.dk.
\end{IEEEbiography}

\begin{IEEEbiography}{Niklas Elmqvist} is a Villum Investigator and professor in the Department of Computer Science at Aarhus University in Aarhus, Denmark.
His research interests include data visualization, human-computer interaction, and human-centered AI.
Elmqvist received the Ph.D.\ degree in 2006 in computer science from Chalmers University of Technology.
He is a Fellow of the IEEE and the ACM.
Contact him at elm@cs.au.dk.
\end{IEEEbiography}

\end{document}